\documentclass[epj,nopacs]{svjour}

\usepackage{amsmath}
\usepackage{graphicx}
\usepackage{widetext}

\newcommand{\eg}{e.g., }

\newcommand{\etas}{\eta_{\rm s}}
\newcommand{\fp}{f_{\rm p}}
\newcommand{\fs}{f_{\rm s}}
\newcommand{\Gp}{G_{\rm p}}
\newcommand{\ie}{i.e., }
\newcommand{\Kp}{K_{\rm p}}
\newcommand{\kT}{k_{\rm B}T}
\newcommand{\lc}{\ell_{\rm c}}
\newcommand{\Le}{L_{\rm e}}
\newcommand{\Lp}{L_{\rm p}}
\newcommand{\Lpp}{\Lambda_{\rm pp}}
\newcommand{\Lss}{\Lambda_{\rm ss}}
\newcommand{\Lps}{\Lambda_{\rm ps}}
\newcommand{\Lsp}{\Lambda_{\rm sp}}
\newcommand{\MSD}{{\langle(\Delta h)^2(t)\rangle}}
\newcommand{\MSDw}{{\langle(\Delta h)^2(\omega)\rangle}}
\newcommand{\pd}{\partial}

\newcommand{\rmd}{{\rm d}}
\newcommand{\tauc}{\tau_{\rm c}}
\newcommand{\taue}{\tau_{\rm e}}
\newcommand{\tauL}{\tau_{\phi L}}
\newcommand{\taup}{\tau_{\phi{\rm p}}}
\newcommand{\tauR}{\tau_{\rm R}}
\newcommand{\taurep}{\tau_{\rm rep}}
\newcommand{\tenG}{{\boldsymbol{\cal G}}}

\newcommand{\veck}{{\bf k}}
\newcommand{\vecq}{{\bf q}}
\newcommand{\vecr}{{\bf r}}
\newcommand{\vecrho}{\boldsymbol{\rho}}

\newcommand{\vp}{v_{\rm p}}
\newcommand{\vs}{v_{\rm s}}

\begin{document}

\title{Membrane undulations in a structured fluid: Universal dynamics
  at intermediate length and time scales}

\titlerunning{Membrane dynamics in a structured fluid}

\author{Rony Granek\inst{1} \and Haim Diamant\inst{2}}

\authorrunning{R. Granek and H. Diamant}

\institute{The Stella and Avram Goren-Goldstein Department of
Biotechnology Engineering, Ben-Gurion University of The Negev,
Beer Sheva 84105, Israel \and
Raymond \& Beverly Sackler School of Chemistry,
Tel Aviv University, Tel Aviv 69978, Israel}

\date{Received: date / Revised version: date}

\abstract{The dynamics of membrane undulations inside a viscous solvent is
governed by distinctive, anomalous, power laws.
Inside a viscoelastic continuous medium these universal behaviors are
modified by the specific bulk viscoelastic spectrum. Yet, in
structured fluids the continuum limit is reached only beyond a
characteristic correlation length. We study the
crossover to this asymptotic bulk dynamics. The analysis
relies on a recent generalization of the hydrodynamic interaction in
structured fluids, which shows a slow spatial decay of the interaction
toward the bulk limit.
For membranes which are weakly coupled to the structured medium we
find a wide crossover regime characterized by different, universal,
dynamic power laws.
We discuss various systems for which this behavior is relevant, and
delineate the time regime over which it may be observed.
%
} 

\maketitle






\section{Introduction}
\label{sec_intro}

The main building block of biological membranes is a flexible fluid
bilayer of phospholipid molecules \cite{AlbertsBook}. Both the
equilibrium and dynamic properties of this system have been vastly
investigated (see, \eg Refs.~\cite{Seifert1997,SafranBook}). As
regards the linear response to out-of-plane forces, and the
corresponding dynamics of fluctuations, most theoretical studies have
considered membranes surrounded by a simple viscous solvent (\ie
water)
\cite{Brochard1975,Seifert1993,Zilman1996,Zilman2002,Watson2011,Bingham2015}. The
strong fluctuations of tensionless membranes, along with the
instantaneous response of the viscous solvent, result in anomalous
dynamic exponents. The relaxation rate is much slower than that of an
ordinary tension-dominated surface, scaling with wavenumber $k$ as
$\Omega(k)\sim k^3$ rather than $\sim k$. The mean-square displacement
(MSD) of a membrane segment in the transverse direction is
subdiffusive, scaling with time $t$ as $\MSD \sim t^{2/3}$ \cite{Granek1997}, in between
the confined fluctuations ($\sim t^0$) of elastic surfaces, on the one
hand, and normal diffusion ($\sim t$), on the other. The dynamic
structure factor of membrane fluctuations follows a
stretched-exponential decay, $S(q,t)\sim \exp[-(\Gamma_q t)^{2/3}]$,
instead of the ordinary exponential decay in the case of normal
diffusion \cite{Zilman1996}.

Biological membranes, in particular, the plasma membrane of eukaryotic
cells or the inner membrane of bacteria, however, are in contact with
more complex media, such as the cytoskeleton or extra-cellular
matrix. Artificial, self-assembled polymer-membrane complexes have
been also thoroughly investigated
\cite{Auvray1997,Sackmann2000,Sackmann2005,Richter2002,polymer-membrane-DLS,Tsai2011}. The
theory was extended, therefore, to membranes embedded in a
viscoelastic medium \cite{Granek2011}.  In that work the medium was
taken to be a structureless continuum, characterized by a complex
frequency-dependent shear modulus $G(\omega)$. The specific frequency
dependence of $G(\omega)$ modifies the characteristic exponents of
membrane dynamics into specific, medium-dependent ones.

Until recently, the bulk behavior of a continuous medium, as captured
by $G(\omega)$, was thought to hold over distances larger than the
medium's static correlation length $\xi$ (\eg the mesh size of a
polymer network). One of the first derivations of such ``hydrodynamic
screening'' for semidilute flexible polymer solutions was that of
Freed and Edwards, who found that the hydrodynamic interaction within
the network decays as $\exp(-r/\xi)/r$ at distances $r\gg \xi$,
replacing the $1/r$ Oseen interaction in a viscous fluid
\cite{DoiEdwardsBook}. It has recently been discovered, however, that
this description of hydrodynamic screening in structured fluids is
lacking. The bulk behavior has been demonstrated, both experimentally
and theoretically, to set in beyond a larger dynamic crossover
distance \cite{Sonn-Segev2014a,Sonn-Segev2014b,Diamant2015},
\begin{equation}
  \lc(\omega) = \xi\,[\eta(\omega)/\etas]^{1/2},
\label{lc}
\end{equation}
where $\eta(\omega)=G(\omega)/(i\omega)$, and $\etas$ is the shear
viscosity of the solvent.
Usually, over most of the relevant frequency range, one has
$\eta(\omega)\gg\etas$, implying $\lc\gg\xi$.
This opens up an intermediate spatio-temporal regime,
$\xi<r<\lc(\omega)$, over which the dynamics of the medium is
qualitatively different from its bulk behavior.

The purpose of the present work is to investigate the consequences of
the distinct behavior within this intermediate regime for the dynamics
of a membrane embedded in the structured fluid. To account for the
dynamics of the fluid beyond its bulk behavior, we employ the
two-fluid model of polymer networks
\cite{deGennes1976,Doi1992,Milner1993,Levine2001a}.

In Sec.~\ref{sec_model} we present the model system under
study. Subsequently, we begin the analysis in Sec.~\ref{sec_scaling}
with a simple scaling argument, which qualitatively accounts for the
basic effects (including power laws) to be derived in the sections
that follow. The detailed analysis is divided into two stages. We
first derive in Sec.~\ref{sec_kernel} the hydrodynamic interaction
kernels, coupling the membrane with the two constituents of the
surrounding fluid (polymer and solvent). The boundary conditions at
the membrane surface define two limiting cases for the strength of the
membrane--fluid coupling: (a) weak coupling (Sec.~\ref{sec_weak}),
where the membrane is in contact primarily with the solvent, and the
network is affected indirectly, through its coupling to the solvent;
(b) strong coupling (Sec.~\ref{sec_strong}), where both solvent and
network move together with the membrane. We then study in
Sec.~\ref{sec_dynamics} the consequences for the dynamics of membrane
undulations\,---\,in particular, the transverse MSD of membrane
segments (Sec.~\ref{sec_MSD}). We examine the practical relevance of
the general results for two examples of a structured fluid: a
semidilute solution of flexible (Sec.~\ref{sec_flexible}) and
semiflexible (Sec.~\ref{sec_semiflexible}) polymers. Finally, we
discuss in Sec.~\ref{sec_discuss} the various findings, their
limitations and implications.

\section{Model}
\label{sec_model}

\begin{figure}[!t]
  \centerline{
\includegraphics[width=0.47\textwidth]{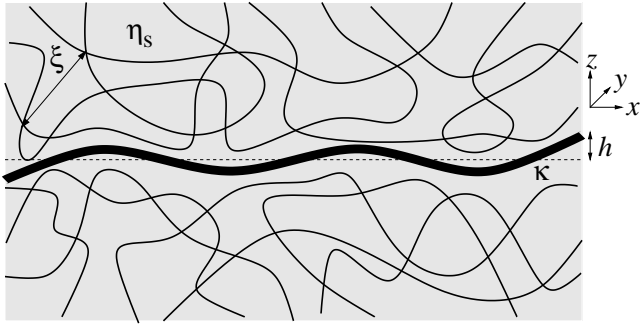}}
\caption{Schematic illustration of the system and its parameters.}
\label{fig_scheme}
\end{figure}

The system is schematically depicted in Fig.~\ref{fig_scheme}.  A
tensionless membrane of bending rigidity $\kappa$ is embedded in a
medium made of a semidilute polymer network of mesh size $\xi$ inside
a solvent of viscosity $\etas$. We neglect the more detailed
inner-membrane dynamics
\cite{Seifert1993,Watson2011,Bingham2015,Fournier}, which takes place
on nanometric length scales.  We use the spatial coordinates
$\vecr=(\vecrho,z)$, where $\vecrho$ is a two-dimensional (2D)
position vector on the $xy$ plane.  The membrane lies on average on
the $xy$ plane and its out-of-plane configuration is parametrized by
the height function $h(\vecrho,t)$.

For the medium we use the two-fluid model, which is a well studied
model of polymer networks
\cite{deGennes1976,Doi1992,Milner1993,Levine2001a,Diamant2015}. The
model accounts for the response of the polymer network via
viscoelastic shear and compression moduli, $\Gp(\omega)$ and
$\Kp(\omega)$, and for its coupling to the solvent via mutual friction
with friction coefficient $\Gamma(\omega)$. The model yields the bulk
shear viscosity as $\eta(\omega)=\Gp(\omega)/(i\omega) + \etas$. It
also produces two characteristic lengths. The first, emerging from the
shear response, is given by $\xi=[\Gp\etas/(i\omega\Gamma\eta)]^{1/2}$
and is identified with the mesh size.\footnote{The apparent frequency
  dependence of $\xi$ is negligible so long as $\Gp(\omega)\gg\etas\omega$,
  \ie for frequencies that are not too high.} The second is related to
the compression response,
$\lambda=[(4\Gp/3+\Kp)/(i\omega\Gamma)]^{1/2} = \xi [2(\eta/\etas)
  (1-\sigma)/(1-2\sigma)]^{1/2}$, where $\sigma$ is the network's
Poisson ratio. The inequality $\lambda/\xi>1$ holds always, turning in
the limit of an incompressible network ($\sigma\rightarrow 1/2$) into
$\lambda/\xi\rightarrow\infty$. From now on we are going to use the
emergent parameters $(\eta(\omega),\xi,\lambda)$, instead of
$(\Gp,\Kp,\Gamma)$.

Because of the two components in the two-fluid model, the hydrodynamic
interactions are described by four tensorial kernels, $\tenG_{\rm
  pp}$, $\tenG_{\rm ss}$, $\tenG_{\rm ps}$, and $\tenG_{\rm sp}$.
They correspond to the velocity response of each of the two
components, polymer or solvent, to a force exerted on either the same
or the other component\,---\,polymer-polymer, solvent-solvent,
polymer-solvent, and solvent-polymer. Due to Onsager's reciprocal
relations, $\tenG_{\rm ps}=\tenG_{\rm sp}$. The three tensors were
calculated in Ref.~\cite{Levine2001a}. Of particular interest here is
the solvent-solvent kernel. Asymptotically, for $r\gg\xi$, it is given
by
\begin{equation}
  \tenG_{ss}(\vecr) \simeq \frac{1}{8\pi\eta(\omega) r} ({\bf 1} +
       {\hat r}\otimes{\hat r}) -\frac{\xi^2}{4\pi\etas r^3} ({\bf 1}
       - 3{\hat r}\otimes{\hat r}).
\label{GssAsymp}
\end{equation}
This modified hydrodynamic interaction tensor, coupling two points
located within the solvent, shows the crossover between the
intermediate and bulk regimes at the distance $\lc(\omega)$ as given
by Eq.~(\ref{lc}). The first term, decaying as $1/r$ and dominating at
$r\gg\lc$, is the usual Oseen tensor. It governs the interaction at
long distances and is controlled by the bulk viscoelasticity. The
second term, decaying as $1/r^3$, dominates at $\xi \ll r\ll \lc$ and
depends on the much lower solvent viscosity. (A third,
solvent-dominated regime at $r\ll\xi$ is missing from this asymptotic
expression.) For actin networks, for example, the intermediate
behavior was observed at distances of a few microns
\cite{Sonn-Segev2014a,Sonn-Segev2014b}.

It is important to note that the asymptotic two terms in
Eq.~(\ref{GssAsymp}) reflect two conservation laws and, as such, do
not depend on any specific model such as the two-fluid one
\cite{Diamant2007,Sonn-Segev2014a,Diamant2015}. The first term arises
from momentum conservation of the entire medium; it describes the flow
velocity due to the momentum monopole created by the force. The second
term comes from mass conservation of the solvent as it flows past the
network; it describes the flow due to the effective mass dipole
created by the force over the mesh size $\xi$.

Because of its dipolar shape, the $1/r^3$ term in Eq.~(\ref{GssAsymp})
vanishes identically under angular averaging. Thus, it does not affect
properties such as the medium's dynamic structure factor or the
pre-averaged correlation between two polymer segments. One is then
left with the usual asymptotic (angle-averaged, diagonal) Oseen
interaction $1/(6\pi\eta r)$ \cite{DoiEdwardsBook}. In the present
case, however, the nearly planar membrane breaks the isotropy of the
system, and the $1/r^3$ term contributes to the hydrodynamic
interaction between membrane segments. As we shall see, this may
change the crossover from solvent- to bulk-dominated dynamics of
membrane undulations.

We tackle the problem from two different directions, both yielding
identical results. In the first route, presented in the main text, we
base the analysis on the 2D hydrodynamic kernel coupling two membrane
segments. We obtain this kernel from the 3D kernels of the two-fluid
model for two extreme situations. In one the polymer network is
depleted from the membrane (weak-coupling limit), and in the other the
polymer network is strongly adsorbed or anchored to the membrane
(strong-coupling limit). In the second route, presented in the
Appendix, we study the membrane dynamics by solving a hydrodynamic
boundary value problem.

\section{Scaling approach}
\label{sec_scaling}

We commence by presenting a simple scaling argument for the appearance
of an intermediate dynamical regime of membrane undulations, assuming
that the membrane interacts primarily with the solvent.

Consider two points on the membrane, separated by a projected distance
$\rho$. Hydrodynamic interaction makes the transverse velocity of the
membrane at one point, $\pd h/\pd t$ respond to the transverse force
density (per unit area) $f$ exerted on it at the other point, as $\pd
h/\pd t=\Lambda(\rho) f$. According to Eq.~(\ref{GssAsymp}), for $\xi
\ll \rho\ll \lc$, the dominant hydrodynamic interaction obeys
$\Lambda(\rho)\sim \xi^2/(\eta_s \rho^3)$. Its 2D Fourier transform is
$\Lambda(k)\sim \xi^2 k/\eta_s$. Substituting the force density due to
bending, $f=\kappa\nabla^4 h$, we obtain the relaxation rate of
undulation mode $\veck$ as
\begin{equation}
  \Omega(k)=\Lambda(k)\kappa k^4\sim \kappa \xi^2 k^5/\eta_s.
\end{equation}
Note the higher power law, $\sim k^5$, compared to the conventional
$\sim k^3$ law \cite{Brochard1975}. Let us assume, for the sake of the
scaling argument, that the membrane size $\ell$ is in the range $\xi \ll
\ell\ll \ell_c$. Thus, the longest undulation relaxation time obeys
\begin{equation}
  \tau(\ell)\sim [\Omega(k=\pi/\ell)]^{-1} \sim \eta_s \ell^5/(\kappa\xi^2).
\end{equation}

We now assume the following scaling hypothesis for the transverse MSD
of membrane segments,
\begin{equation}
  \MSD \equiv \langle \left(h(t)-h(0)\right)^2 \rangle =
  \langle h^2\rangle_{\text{eq}}
  {\cal U}(t/\tau(\ell)),
\end{equation}
where ${\cal U}(x)$ is a scaling function, and $\langle
h^2\rangle_{\text{eq}}$ is the equilibrium mean-square undulation. As
is well known \cite{Seifert1997,SafranBook}, $\langle
h^2\rangle_{\text{eq}}\sim (\kT/\kappa) \ell^2$. Since the MSD should be
independent of $\ell$ for $t\ll \tau(\ell)$, it follows that the scaling
function must behave as ${\cal U}(x)\sim x^{2/5}$ for $x\ll 1$,
leading to
\begin{equation}
  \MSD \sim (\kT/\kappa)\xi^2 (t/\tau_{\xi})^{2/5},
\label{MSDscaling}
\end{equation}
where $\tau_{\xi}\simeq \eta_s\xi^3/\kappa$ is the undulation
relaxation time of a membrane patch of size $\xi$. Thus, $\MSD\sim
t^{2/5}$, with a new anomalous diffusion exponent, $2/5$, replacing the
conventional $2/3$ exponent \cite{Zilman1996}.

\vspace{0.5cm}
\section{Hydrodynamic interaction kernel}
\label{sec_kernel}

As the first step in the detailed analysis, we calculate the 2D
hydrodynamic-interaction kernel, $\Lambda(\vecrho,t)$, correlating two
points on the membrane in space and time. As the membrane lies on
average on the $xy$ plane and fluctuates in the $z$ direction,
$\Lambda(\vecrho,t)$ is generally obtained from a given 3D
hydrodynamic kernel of the medium, ${\cal G}_{ij}(\vecr,t)$, as
\begin{equation}
  \Lambda(\vecrho,t) = {\cal G}_{zz}(\vecr=(\vecrho,0),t).
\end{equation}
We use throughout this article Fourier transforms in 2D space,
$\vecrho\rightarrow\veck$ and in 3D space, $\vecr\rightarrow\vecq$, as
well as Fourier-Laplace transforms in time, $t\rightarrow\omega$. The
transformed 2D kernel is
\begin{equation}
  \Lambda(\veck,\omega) = \frac{1}{2\pi}\int_{-\infty}^\infty \rmd q_z
  {\cal G}_{zz}(\vecq=(\veck,q_z),\omega).
\label{LambdaG}
\end{equation}
If the system is isotropic within the $xy$ plane, then
$\Lambda(\veck,\omega)=\Lambda(k,\omega)$.

For the two-fluid medium the situation is slightly more complicated.
Consider a force density $f$ in the $z$ direction, exerted by the
membrane on the surrounding medium. Let us assume that a fraction
$\alpha$ of this force is exerted on the polymer network, $\fp=\alpha
f$, and the rest is exerted on the solvent, $\fs=(1-\alpha) f$. The
transverse velocities of the two components at $z=0$ are then,
\begin{eqnarray}
  \vp(\veck,\omega) &=& \left[\alpha\Lpp(\veck,\omega) +
  (1-\alpha)\Lps(\veck,\omega)\right] f,
\nonumber\\
  \vs(\veck,\omega) &=& \left[\alpha\Lsp(\veck,\omega) +
  (1-\alpha)\Lss(\veck,\omega)\right] f,
\label{partition}
\end{eqnarray}
where $\{\Lambda_{ij}\}$ are 2D hydrodynamic kernels, and $\Lps=\Lsp$
from Onsager's reciprocity.  To obtain $\{\Lambda_{ij}\}$ we should
substitute in Eq.~(\ref{LambdaG}) the 3D kernels of the two-fluid
model, $\tenG_{\rm pp}$, $\tenG_{\rm ss}$, and $\tenG_{\rm
  ps}=\tenG_{\rm sp}$ \cite{Levine2001a}. Performing the
integration in Eq.~(\ref{LambdaG}) gives,\\
\begin{widetext}
\begin{subequations}
\label{Lambdas}
\begin{eqnarray}
  \Lpp(k,\omega) &=& \frac{1}{4\eta(\omega)k} \left[
    1 - 2\xi^2k^2  \frac{\eta}{\etas} \left( 1 + \frac{\etas}{\eta} -
  \frac{1}{1-\etas/\eta} \frac{\sqrt{1+\lambda^2k^2}}{\lambda k}
  + \frac{\etas/\eta}{\eta/\etas-1} \frac{\xi k}{\sqrt{1+\xi^2k^2}}
  \right) \right],
\label{Lambdapp}\\
  \Lss(k,\omega) &=& \frac{1}{4\eta(\omega) k} \left[1 +
    2\xi^2k^2\left(\frac{\eta}{\etas}-1 \right) \left(1 - \frac{\xi
      k}{\sqrt{1+\xi^2k^2}}\right) \right],
\label{Lambdass}\\
  \Lps(k,\omega) &=& \frac{1}{4\eta(\omega) k} \left[1 -
    2\xi^2k^2 \left(1 - \frac{\xi k}{\sqrt{1+\xi^2 k^2}} \right) \right].
\label{Lambdaps}
\end{eqnarray}
\end{subequations}
\end{widetext}

Over large distances the two components, polymer and solvent, move
collectively as a single continuum \cite{Diamant2015}. Accordingly, in
the limit $k\ll\lc^{-1}<\xi^{-1}$, Eqs.~(\ref{Lambdas}) give
$\Lpp\simeq\Lss\simeq\Lps\simeq 1/(4\eta(\omega)k)$. Examining the
solvent-solvent kernel in more detail, we find the following limiting
behaviors:
\begin{equation}
  \Lss(k,\omega) \simeq \left\{
  \begin{array}{ll}
  \frac{1}{4\eta(\omega)k},\ \ \ & k\ll\lc^{-1} \\
  \frac{\xi^2k}{2\etas}, & \lc^{-1}\ll k\ll \xi^{-1} \\
  \frac{1}{4\etas k}, & k\gg\xi^{-1},
  \end{array} \right.
\label{regimes}
\end{equation}
displaying all three regimes: bulk, intermediate, and
solvent-dominated. The expression in the intermediate regime agrees
with the one used heuristically in Sec.~\ref{sec_scaling}.

To apply the hydrodynamic kernels of Eqs.~(\ref{Lambdas}) we need to
know how the force exerted by the membrane is distributed between the
two components of the surrounding medium. This will determine the
effective kernel, $\bar\Lambda(k,\omega)$, governing membrane
fluctuations. While the membrane is always in close contact with the
solvent, its coupling to the polymer may be of different strengths. In
the following two sub-sections we consider two limiting cases for the
coupling strength.

\subsection{Weak coupling}
\label{sec_weak}

Assume a polymer network that is inert to the membrane, such that only
excluded-volume interactions exist between them. Consider
Eq.~(\ref{partition}) for such a case. The mean fractions of
membrane-polymer and membrane-solvent collisions are assumed to be
$\phi$ and $(1-\phi)$, respectively, where $\phi$ is the polymer
volume fraction. This implies $\alpha=\phi$ in
Eq.~(\ref{partition}). Taking $\phi\ll 1$, and noting that for
realistic parameters $\Lpp\phi\ll\Lps$ and $\Lsp\phi\ll\Lss$, we have
$\vp\simeq\Lps f$ and $\vs\simeq\Lss f$. Since, in addition,
$\Lps\ll\Lss$, this implies $\vp\ll\vs$, \ie the polymer moves much
more slowly than the solvent. One concludes that the solvent moves
with the membrane while the polymer is effectively frozen at some
distance away. Thus, in this weak membrane-polymer coupling we may
take
\begin{equation}
  \bar\Lambda(k,\omega) \simeq \Lss(k,\omega).
\end{equation}

\begin{figure}
  \vspace{0.7cm}
  \centerline{
  \includegraphics[width=0.47\textwidth]{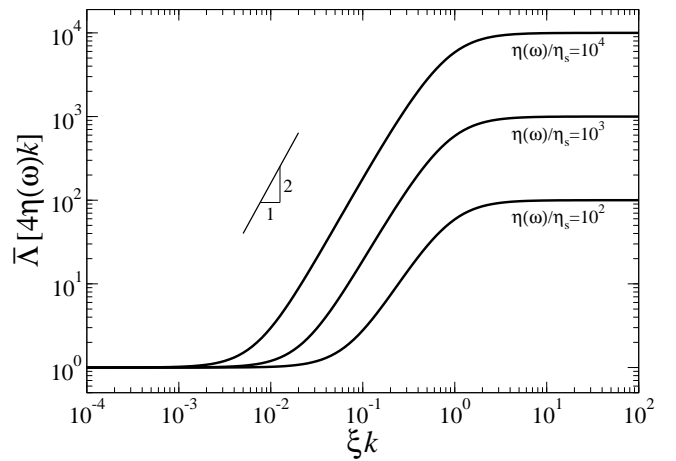}}
\caption{Membrane hydrodynamic kernel in the weak-coupling limit. The
  kernel, normalized by its bulk ($k\rightarrow 0$) expression, is
  plotted against the wavevector, normalized by the inverse mesh size
  of the network, at fixed frequency. In between the bulk limit
  (left-hand side) and the solvent-dominated limit (right-hand side)
  there is an intermediate region where $\bar\Lambda \sim k$. The
  width of this region increases with $[\eta(\omega)/\etas]^{1/2}$
  (curves from right to left).}
\label{fig_LambdaFree}
\end{figure}

In the alternative boundary-value formulation (see the Appendix) we
find that this limit is equivalent to assuming a no-slip boundary
condition for the solvent and free (zero-stress) boundary condition
for the polymer network. We will comment further on the physical
relevance of this case in Sec.~\ref{sec_discuss}. Note that the kernel
in this limit is independent of $\lambda$, i.e., of network
compressibility (see Eq.~(\ref{Lambdass})). This is because the
membrane applies compressive stress exclusively on the solvent, and
the network is displaced only due to the frictional coupling with the
solvent. Figure \ref{fig_LambdaFree} shows the weak-coupling kernel as
a function of wavevector, exhibiting the solvent-dominated,
intermediate, and bulk regimes. The width of the intermediate regime
is proportional to $[\eta(\omega)/\etas]^{1/2}$.

\subsection{Strong coupling}
\label{sec_strong}

The other limit is that of strong coupling of the membrane to both
solvent and polymer. This will be the case when the network is
anchored, or strongly adsorbed, to the membrane.  In this case we find
the fraction $\alpha$ in Eq.~(\ref{Lambdas}) by demanding that the two
components have the same velocity at every point on the membrane at
all times,
$\alpha\Lpp+(1-\alpha)\Lps=\alpha\Lsp+(1-\alpha)\Lss=\bar\Lambda$. This
gives
\begin{eqnarray}
\label{alpha}
   \alpha(k,\omega) = &&\left(1 - \frac{\etas}{\eta(\omega)} \right)
   \lambda k \times \\
   &&\frac{1 - \xi k(\sqrt{1+\xi^2k^2} - \xi k)}
   {(1+\xi^2k^2)\sqrt{1+\lambda^2k^2} - \xi\lambda k^2\sqrt{1+\xi^2k^2}}.
   \nonumber
\end{eqnarray}
Using Eqs.~(\ref{Lambdass}) and (\ref{Lambdaps}), we obtain
\begin{eqnarray}
  \bar\Lambda(k,\omega)=
  &&\frac{1}{4\eta(\omega) k} \left[1 +
    2\xi^2k^2\left(\frac{\eta(\omega)}{\etas}\left(1-\alpha\right)-1
    \right)\times\right. \nonumber\\
    &&\left.\left(1 - \frac{\xi k}{\sqrt{1+\xi^2k^2}}\right) \right].
\label{Lambdastrong}
\end{eqnarray}
An identical kernel to Eq.~(\ref{Lambdastrong}) is obtained from the
boundary-value formalism by imposing no-slip boundary conditions on
both solvent and polymer (see the Appendix).


A particularly simple limit is found for an incompressible polymer
network, where Eqs.~(\ref{alpha}) and (\ref{Lambdastrong}) reduce to
\begin{eqnarray}
  \alpha(k,\omega)
  &\xrightarrow{\lambda\rightarrow\infty}&
  1 - \frac{\etas}{\eta(\omega)}
  = \frac{\Gp(\omega)}{\Gp(\omega)+i\omega\etas}, \nonumber\\
  \bar\Lambda(k,\omega) &\xrightarrow{\lambda\rightarrow\infty}&
  \frac{1}{4\eta(\omega) k}.
\label{highcomplimit}
\end{eqnarray}
Thus, in this limit the force is everywhere distributed according to
the relative resistance of the components to shear, and uniform
viscoelasticity applies essentially at all wavelengths. This result is
a consequence of the nearly planar membrane geometry, where the stress
applied by the membrane is in the purely normal direcion, i.e., has
only a $zz$ component. (This is not true, for example, in the case of
a sphere moving through the two-fluid medium \cite{Diamant2015}.)
Therefore, if both solvent and network are incompressible, they are
bound to be displaced together by the $zz$ stress, implying that the
relevant viscosity everywhere is the collective one, $\eta(\omega)$.

The behavior just described suppresses the intermediate regime in the
strong-coupling case even for compressible networks. To see this, we
refine the criterion for Eq.~(\ref{highcomplimit}) to $\lambda\gg
\xi[\eta(\omega)/\eta_s]^{1/2}\sim\lc$, corresponding to Poisson ratio
$|1/2-\sigma|\ll 1$. (This usually does not hold for polymeric
networks.)
In the large-wavelength regime, $\lambda k\ll 1$, where the polymer
network is effectively compressible, we have $\alpha\ll 1$ [see
  Eq.~(\ref{alpha})].
This implies that $\bar\Lambda(k,\omega)\simeq \Lss(k,\omega)$; yet,
for such small $k$, $\Lss$ already behaves as the bulk kernel,
$\Lss(k,\omega)\simeq 1/(4\eta(\omega) k)$, and the intermediate
regime will not be observed. For $\lambda\sim\lc$, we expect some
deviation from the limit of Eq.~(\ref{highcomplimit}) around $\lambda
k\sim\lc k\sim 1$.

\begin{figure}
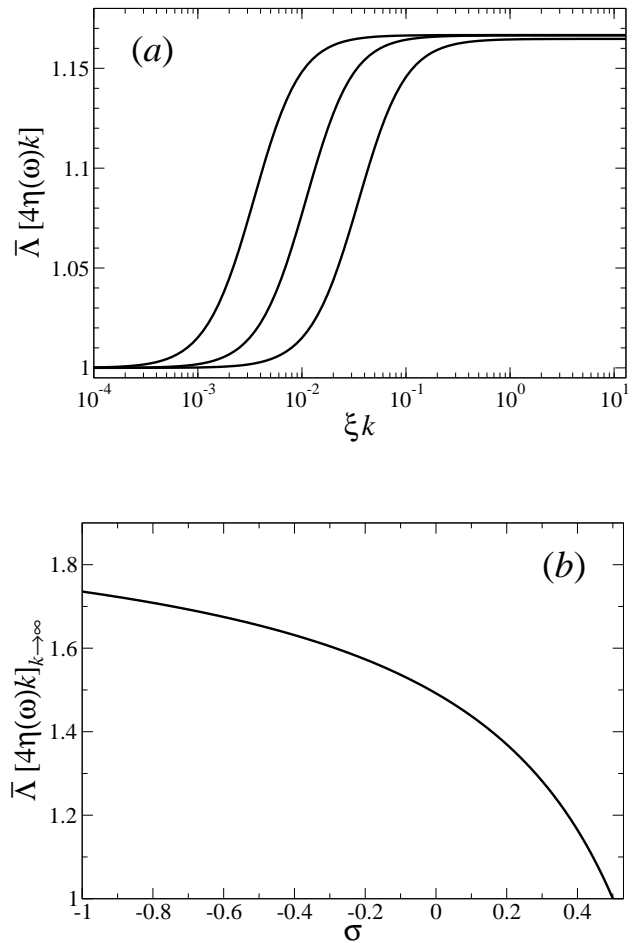

  \vspace{0.4cm}
  \centerline{
\includegraphics[width=0.45\textwidth]{fig3a.eps}}
\vspace{1cm}
\centerline{\includegraphics[width=0.45\textwidth]{fig3b.eps}}
\caption{(a) Membrane hydrodynamic kernel in the strong-coupling
  limit. The kernel, normalized by its bulk expression, is plotted
  against the wavevector, normalized by the inverse mesh size of the
  network, at fixed frequency, for Poisson ratio $0.4$. The three
  curves, from right to left, correspond to $\eta(\omega)/\etas=10^2$,
  $10^3$, and $10^4$. (b) Large-wavevector limit of the kernel as a
  function of Poisson ratio, for $\eta(\omega)/\etas=10^2$. Note the
  narrow range of values (panel (a)), which completely disappears for
  $\sigma=1/2$ (panel (b)).}
\label{fig_LambdaStick}
\end{figure}

These observations are confirmed in Fig.~\ref{fig_LambdaStick}. Panel
({\it a}) shows the strong-coupling kernel as a function of $k$, for
three different values of $\eta(\omega)/\etas$ and Poisson ratio
$\sigma=0.4$. Note how close the kernel remains to its bulk limit, due
to the response of the medium to compression. For the same reason the
kernel exhibits neither the solvent-dominated behavior for $\xi k\gg
1$ nor the intermediate region. The range of values narrows down to
zero as the limit of incompressible network ($\sigma=1/2$) is
approached. Figure~\ref{fig_LambdaStick}({\it b}) presents the
large-$k$ limit of the kernel as a function of Poisson ratio. The
ratio between this small-scale limit and the bulk one remains of order
1 over the full range of Poisson ratios.

It is easy to generalize these results to the case of two different
viscoelastic media ('1' and '2') surrounding the membrane, as often
occurs for biological membranes (\eg the cytoskeleton and the
extracellular matrix). The half space (hf) interaction kernel for each
side is twice that of the full space kernel, $\bar\Lambda_i^{\rm
  (hf)}=2\bar\Lambda_i$, $i=1,2$. Distributing the membrane force
between the two sides with fractions $\beta$ and $1-\beta$, and
requiring that the membrane velocity be unique, we have $\bar\Lambda=
\bar\Lambda_1^{\rm (hf)}\beta=\bar\Lambda_2^{\rm (hf)}(1-\beta)$.
This leads to $\beta=\bar\Lambda_2^{\rm (hf)}/(\bar\Lambda_1^{\rm
  (hf)}+\bar\Lambda_2^{\rm (hf)})$, and
\begin{equation}
  \bar\Lambda=\frac{2\bar\Lambda_1\bar\Lambda_2}
  {\bar\Lambda_1+\bar\Lambda_2}.
\end{equation}
In the regimes where bulk viscoelasticity dominates in both media,
this amounts to replacing $\eta(\omega)$ by
$\eta_{\text{eff}}=[\eta_1(\omega)+\eta_2(\omega)]/2$, which is a
known result. Note however that the two media are not necessarily
found in this regime together.

\section{Membrane dynamics in a structured fluid}
\label{sec_dynamics}

Given the hydrodynamic interaction kernel, as calculated in
Sec.~\ref{sec_kernel}, we may write down a generalized Langevin
equation of motion for the membrane's displacement field
$h(\vecrho,t)$ assuming small deformations \cite{Granek2011},
\begin{eqnarray}
  \pd_t h(\vecrho,t) = && -\int_0^t \rmd t'\int
  \rmd^2\rho'\bar\Lambda(|\vecrho-\vecrho'|, t-t')
  \,\kappa \nabla^4_{\rho'}h(\vecrho',t') \nonumber\\
  && + \zeta(\vecrho,t).
\label{Langevin-rho}
\end{eqnarray}
Here $\zeta(\vecrho,t)$ is a thermal colored noise, obeying
the \\ fluctuation-dissipation theorem,
\begin{equation}
  \langle\zeta(\vecrho,t)\zeta(\vecrho',t')\rangle
  = \kT \bar\Lambda(|\vec\rho-\vec{\rho '}|,|t-t'|).
\end{equation}

Applying to Eq.\ (\ref{Langevin-rho}) a Fourier transform in
$\vecrho$ and a Fourier-Laplace transform in $t$ yields
\begin{equation}
  i\omega h(\veck,\omega) - h(\veck,t=0)
  = -\bar\Lambda(k,\omega) \kappa k^4
  h(\veck,\omega) + \zeta(\veck,\omega).
\label{Langevin-k-omega}
\end{equation}
Solving Eq.\ (\ref{Langevin-k-omega}) for $h$, we find
\begin{equation}
  h(\veck,\omega) = \frac{h(\veck,t=0) + \zeta(\veck,\omega)}
  {i\omega + \Omega(k,\omega)},
\label{h-k-omega}
\end{equation}
with the generalized relaxation rate,
\begin{equation}
  \Omega(k,\omega) = \bar\Lambda(k,\omega)
  \kappa k^4.
\label{Omega-k-omega}
\end{equation}

Consider now the transverse MSD $\MSD$ and its Fourier-Laplace
transform $\MSDw$. It can be conveniently written as
\cite{Granek2011}
\begin{equation}
  \MSDw = \frac{1}{2\pi^2} \frac{\kT}{\kappa}\int
  \frac{\rmd^2k}{k^4} \left( \frac{1}{i\omega} - \frac{1}
  {i\omega+\Omega(k,\omega)} \right).
\label{Dh-omega1}
\end{equation}
Rearranging the integrand using Eq.\ (\ref{Omega-k-omega}) for
$\Omega$, we have
\begin{equation}
  \MSDw = \frac{1}{\pi} \frac{\kT}{i\omega}
  \int_0^{\infty} \rmd k \frac{k\bar\Lambda(k,\omega)}
  {i\omega + \bar\Lambda(k,\omega) \kappa k^4}.
\label{Dh-omega-2}
\end{equation}

The integral in Eq.~(\ref{Dh-omega-2}) is not easily evaluated
analytically. We turn, therefore, to asymptotic analysis and numerical
integration for several examples. As has been shown in
Sec.~\ref{sec_kernel}, strong membrane-network coupling suppresses the
intermediate dynamics, leaving us with the already known
solvent-dominated \cite{Zilman1996} and bulk \cite{Granek2011}
regimes. Hence, in what follows, we focus on the weak-coupling limit.

\subsection{Transverse MSD of a weakly coupled membrane}
\label{sec_MSD}

Using $\Lss$ of Eq.~(\ref{Lambdass}) in Eq.~(\ref{Dh-omega-2}) and
transforming to a dimensionless wavenumber $\hat{k}= k\xi$, we obtain
\begin{equation}
  \MSDw = A \int_0^{\infty} \rmd\hat{k}
        {\Psi(\hat k,\hat\eta)\over 1 + {\hat\gamma}{\hat
      k}^3 \Psi(\hat k,\hat\eta)},
\label{MSDgeneral}
\end{equation}
where we have defined:
\begin{eqnarray*}
  A &=& \frac{\kT}{4\pi (i\omega)^2\eta(\omega)\xi} \\
  \hat\eta &=& \frac{\eta(\omega)}{\etas} - 1\\
  \hat\gamma &=& {\kappa\over 4i\omega\eta(\omega)\xi^3} \\
  \Psi(\hat k,\hat\eta) &=&  1 + 2\hat\eta{\hat k}^2
  \left(1 - \frac{\hat k}{\sqrt{1+{\hat k}^2}}\right).
\end{eqnarray*}

Studying the dependence of Eq.~(\ref{MSDgeneral}) on $\hat\eta$ and
$\hat\gamma$, we find three asymptotic regimes:

\begin{itemize}

  \item[(i)] {\it Solvent-dominated regime}, which holds at high
    frequencies such that $\hat\gamma\ll 1$ (more precisely,
    $\hat\gamma\ll {\hat\eta}^{-1}$). Here the integral is dominated
    by $\hat k\gg 1$ such that $\Psi \simeq 1 + \hat\eta$.  This leads
    to
\begin{equation}
  \MSDw = A {2\pi \over 3\sqrt{3}}(1+\hat\eta)^{2/3}
        {\hat\gamma}^{-1/3},
\end{equation}
or, in terms of dimensional parameters,
\begin{equation}
  \MSDw = B_1 \frac{\kT}{\kappa^{1/3}\eta_s^{2/3} (i\omega)^{5/3}},
\label{MSDsolvent}
\end{equation}
where $B_1 = 2^{-1/3}3^{-3/2}\simeq 0.153$.  This is the known result
for a membrane embedded in a purely viscous solvent
\cite{Zilman1996,Zilman2002}.

\item[(ii)] {\it Intermediate regime}, holding for intermediate
  frequencies such that ${\hat\eta}^{-1} \ll \hat\gamma \ll
  {\hat\eta}^{3/2}$. Here the integral is dominated by
  ${\hat\eta}^{-1/2} \ll \hat k\ll 1 $, such that $\Psi \simeq
  2\hat\eta{\hat k}^2$. This leads to
\begin{equation}
  \MSDw = A {2^{9/10}\pi\left(\sqrt{5}-1\right)^{1/2}
    \over 5^{5/4}}{\hat\eta}^{2/5}{\hat\gamma}^{-3/5},
\end{equation}
or, assuming that in this regime $\eta(\omega)\gg \eta_s$,
\begin{equation}
  \MSDw = B_2
  \frac{\kT\xi^2}{\kappa^{3/5}\etas^{2/5} (i\omega)^{7/5}},
\label{MSDintermed}
\end{equation}
with $B_2=2^{1/10}\left(\sqrt{5}-1\right)^{1/2}/5^{5/4} \simeq
0.159$. Importantly, the MSD in the intermediate regime is independent
of the bulk viscosity $\eta(\omega)$, leaving the mesh size $\xi$ as
the only network property at play.

\item[(iii)] {\it Bulk regime}, which holds at low frequencies such
  that $\hat\gamma\gg \hat\eta^{3/2}$. Here the integral is dominated
  by $\hat k\ll \hat\eta^{-1/2}$, such that $\Psi \simeq 1$. This
  leads to
\begin{equation}
   \MSDw = A {2\pi \over 3\sqrt{3}}{\hat\gamma}^{-1/3},
\end{equation}
or,
\begin{equation}
  \MSDw = B_1 \frac{\kT}{\kappa^{1/3}
    \eta(\omega)^{2/3} (i\omega)^{5/3}},
\end{equation}
which is the known result for a membrane embedded in a continuous
viscoelastic medium \cite{Granek2011}.

\end{itemize}

In the time domain, the first two regimes become:

\begin{itemize}

\item[(i)] {\it Solvent-dominated regime}, which holds at short times,
  $t\ll \tau_{\xi}$, where $\tau_{\xi}=4\etas\xi^3/\kappa$ is the
  undulation relaxation time of a membrane patch of size $\xi$. In
  this regime,
  \begin{equation}
    \MSD = B_1'\left[\left(\frac{\kT}{\kappa}\right)^{\frac{1}{2}}
      \frac{\kT}{\eta}t\right]^{2/3},
  \end{equation}
where $B_1'=\Gamma[1/3]/(2\pi\, 4^{2/3}) \simeq 0.169$.

\item[(ii)] {\it Intermediate regime}, which holds for intermediate
  times, $\tau_{\xi}\ll t\ll \tauc$. The crossover time
  $\tauc$ is the solution of the equation,
  \begin{equation}
    \tauc=\tau_\xi \left({\eta\left[\omega=\tauc^{-1}
        \right]\over \etas}\right)^{5/2}.
    \label{tauc}
  \end{equation}
  In this new regime,
\begin{equation}
  \MSD = B_2' \frac{\kT\xi^2}{\kappa^{3/5}\etas^{2/5}} t^{2/5},
\end{equation}
where $B_2'=B_2/\Gamma[7/5]\simeq 0.180$.  This result, up to the numerical
prefactor, is the same as the one deduced from the scaling argument of
Sec.~\ref{sec_scaling}.

\end{itemize}

The time dependence in the third, bulk regime, as well as the
crossover time $\tauc$ determining the width of the
intermediate regime, depend on the specific bulk
viscoelasticity. Generally, however, due to the power of $5/2$ in
Eq.~(\ref{tauc}), the crossover time may be very long compared to
$\tau_\xi$, yielding a broad intermediate regime. In the following two
sub-sections we study two specific examples that demonstrate this
behavior.

\subsection{Membrane in a semidilute solution of flexible polymers}
\label{sec_flexible}

Let us take the weakly coupled network to be a semidilute flexible
polymer solution. The network is characterized by the length $L$ of
the polymer chains, the entanglement length $\Le$ (length of chain
segments between entanglements), the mesh size $\xi$, and the
intrinsic time scale $\tau_0=\etas\xi^3/\kT$. The stress relaxation
function is composed of four regimes
\cite{DoiEdwardsBook,RubinsteinBook,Granek1992}: (i) Zimm/Rouse regime
at short times, $t<\taue$, where $\taue\simeq\tau_0$ is the entanglement
time; (ii) ``breathing"-plateau regime at intermediate times,
$\taue<t<\tauR$, where $\tauR=\taue(L/\Le)^2$ is the Rouse time;
(iii) short-time reptation-plateau regime for $\tau_{\rm
  R}<t<\taurep$, where $\taurep=\taue(L/\Le)^3$ is the terminal
(reptation) time; and (iv) long-time reptation regime for $t>\taurep$.

\begin{figure}[tbh]
\vspace{0.8cm}
\centerline{\includegraphics[width=0.45\textwidth]{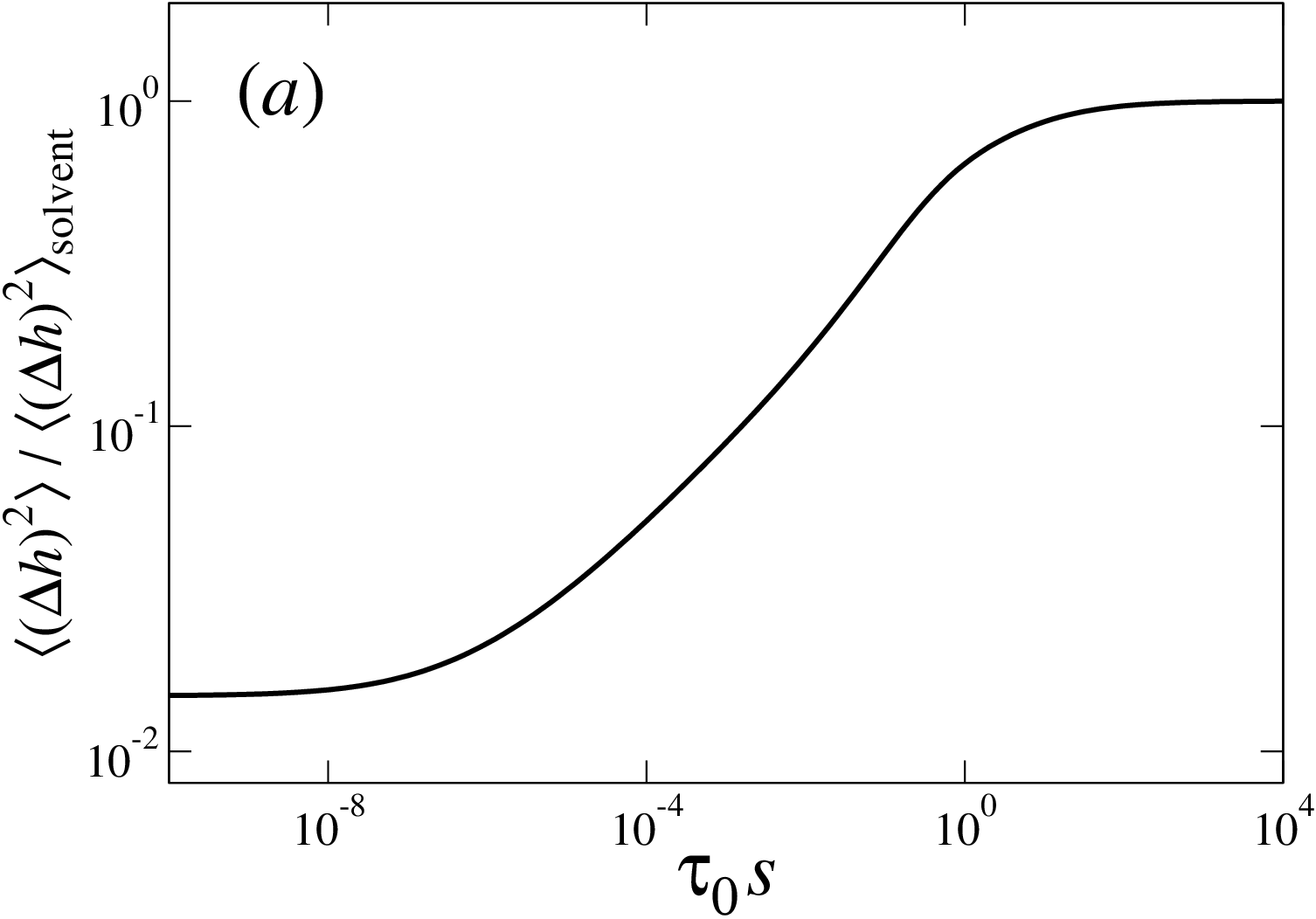}}
\vspace{1cm}
\centerline{\includegraphics[width=0.45\textwidth]{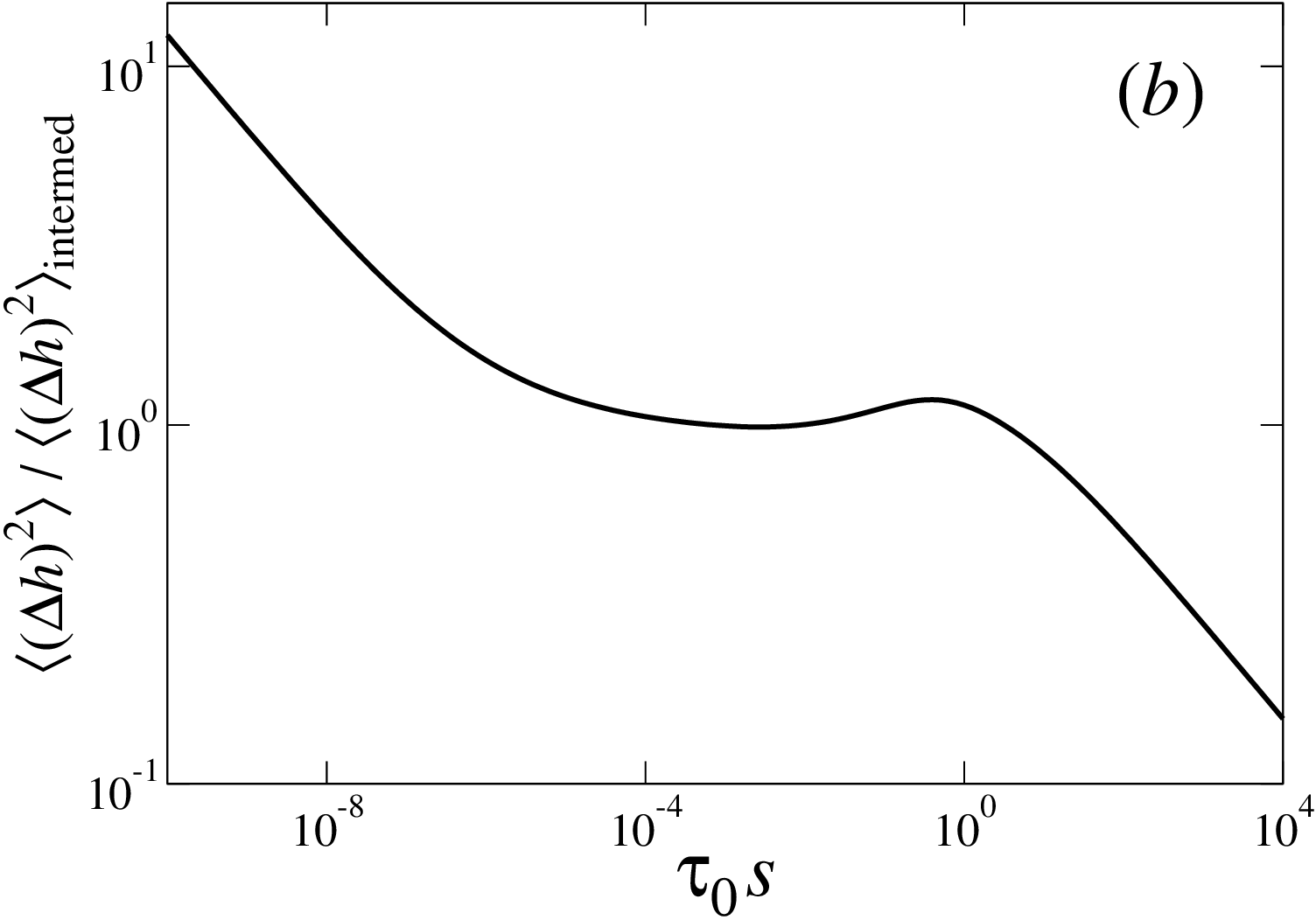}}
\caption{Normalized MSD of a membrane in a flexible polymer network as
  a function of frequency $s=i\omega$. (a) The MSD is normalized by
  its solvent-dominated asymptote, $\langle\Delta h^2\rangle_{\rm
    solvent} \sim s^{-5/3}$ [Eq.~(\ref{MSDsolvent})]. (b) The MSD is
  normalized by its intermediate-regime asymptote, $\langle\Delta
  h^2\rangle_{\rm intermed} \sim s^{-7/5}$
  [Eq.~(\ref{MSDintermed})]. The frequency in both panels is
  normalized by $\tau_0^{-1}=\kT/(\etas\xi^3)$. Parameters:
  $\kappa/\kT=10$, $L/\xi=100$, $\Le/\xi=10$.}
\label{fig_MSDflex}
\end{figure}

Figure~\ref{fig_MSDflex} shows the membrane's transverse MSD for a
strongly entangled case, $L/\Le=10$, as a function of the frequency
$s=i\omega$. These results were obtained by substituting the
viscoelastic shear modulus accounting for the above regimes in
Eq.~(\ref{MSDgeneral}). We clearly see the two asymptotes, $\sim
s^{-5/3}$, at small and large $s$ values, and the intermediate regime,
$\sim s^{-7/5}$, as predicted in Sec.~\ref{sec_MSD}.

The intermediate regime spans 6--7 decades. Indeed, since the short and
intermediate time behaviors have been argued above to be independent
of the network's complex modulus, the only relevant parameter is the
crossover time $\tauc$. A consistent solution of Eq.~(\ref{tauc}) can
be obtained only if the complex modulus is already in the long-time
(small $s$) reptation regime, \ie where the bulk is purely viscous,
having an effective viscosity $\eta(0)\sim\etas(L/\Le)^3$. Solving for
$\tauc$, we find $\tauc\sim\tau_0(L/\Le)^{15/2}$, which is consistent
with the wide frequency range shown in
Fig.~\ref{fig_MSDflex}. Obviously, we do not expect such a wide
intermediate regime in reality; a smaller crossover time, $t^*<\tauc$,
arising from different physics, is expected to precede $\tauc$, as
will be discussed in Sec.~\ref{sec_discuss}.

\subsection{Membrane in a semidilute solution of semiflexible polymers}
\label{sec_semiflexible}

We now turn to the case of a membrane weakly coupled to a semidilute
solution of semiflexible polymers, \eg an entangled F-actin
network. Such a network has another intrinsic length\,---\,the polymer
persistence length $\Lp$. Assuming $\Lp\gg\xi$, we have
$\Le\simeq\xi^{4/5}\Lp^{1/5}$ \cite{Broedersz2014}. The following
hierarchy of time scales emerges \cite{Gittes1998,Morse1998b}:
entanglement time, $\taue=\tau_0(\xi/\Lp)^{1/5}$; two relaxation times
related to filament undulations, $\taup=\tau_0(\Lp/\xi)^{3/5}$ and
$\tauL=\tau_0(L/\Lp)^{7/5}(L/\xi)^{3/5}$; and reptation time,
$\taurep=\tau_0(L/\xi)^3$. For actin networks, typically, $\xi\sim
0.1$~$\mu$m, $\Lp\sim 10$~$\mu$m, and $L\sim 20$~$\mu$m, yielding
$\taue \sim \tau_0 \sim 10^{-4}$~s, $\taup \sim 10^{-3}$~s, $\tauL
\sim 10^{-2}$~s, and $\taurep\sim 10^3$~s. This hierarchy defines the
following regimes for the frequency-dependent response of the network
\cite{Gittes1998,Morse1998b}:
\begin{eqnarray}
\label{Gsemiflex}
  G(\omega) = && i\omega\eta(\omega)
  \simeq b G_0 \times \\
  &&\left\{ \begin{array}{ll}
  (\Lp/\xi)^{5/4} (i\omega\tau_0)^{3/4}, & \omega > \taue^{-1} \\
  (\Lp/\xi)^{7/5}, & \taup^{-1} < \omega < \taue^{-1} \\
  (\Lp/\xi)^{17/10} (i\omega\tau_0)^{1/2}, & \tauL^{-1} < \omega < \taup^{-1} \\
  (\Lp/\xi)^{7/5}(\Lp/L), & \taurep^{-1} < \omega < \tauL^{-1} \\
  (\Lp/\xi)^{22/5}(L/\Lp)^2 (i\omega\tau_0),\ \ \ & \omega < \taurep^{-1}
  \end{array} \right. \nonumber
\end{eqnarray}
where $G_0=\kT/\xi^3$, and $b\sim 0.1$ is a numerical prefactor.

\begin{figure}[tbh]
\vspace{0.8cm}
\centerline{\includegraphics[width=0.45\textwidth]{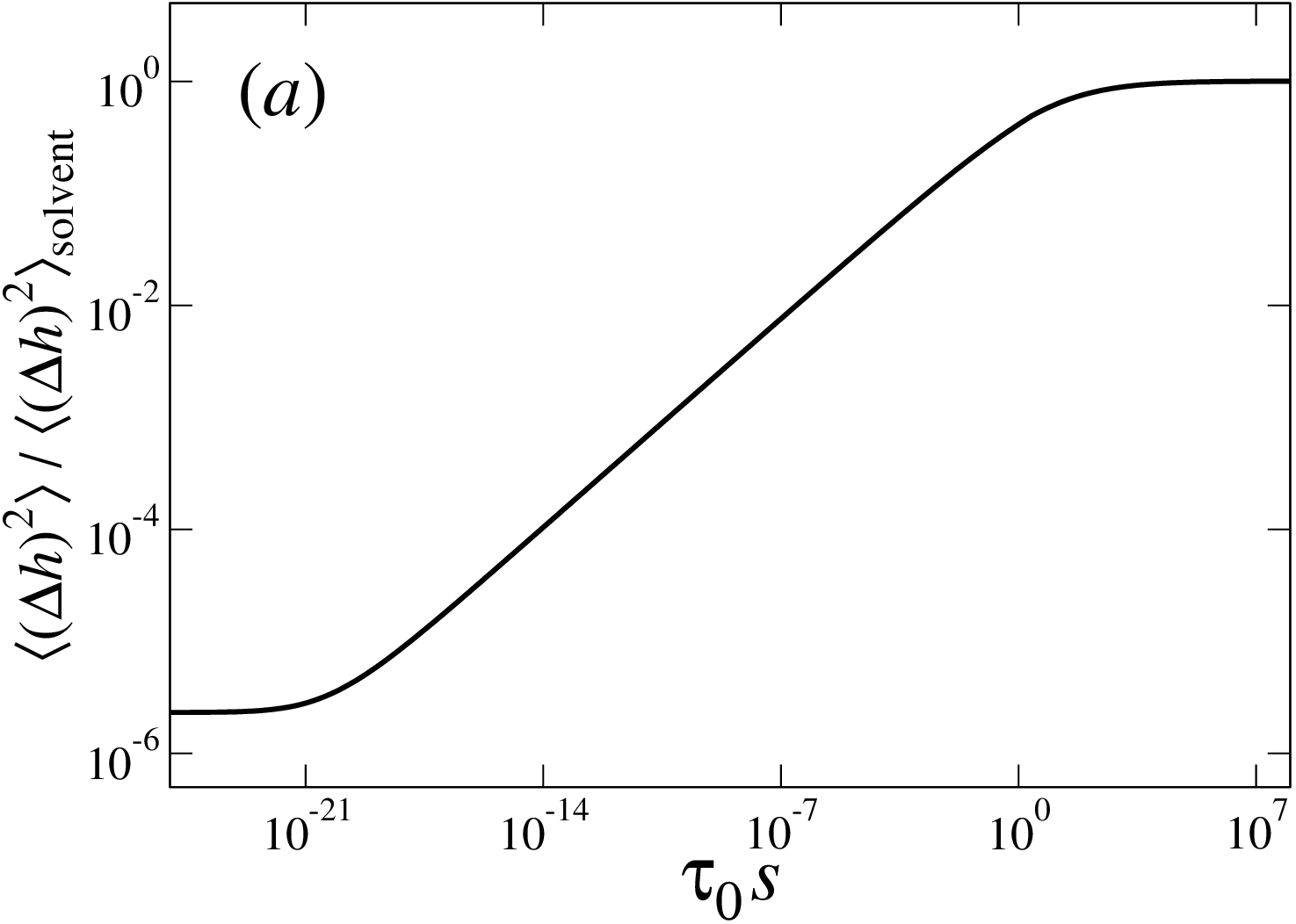}}
\vspace{1cm}
\centerline{\includegraphics[width=0.45\textwidth]{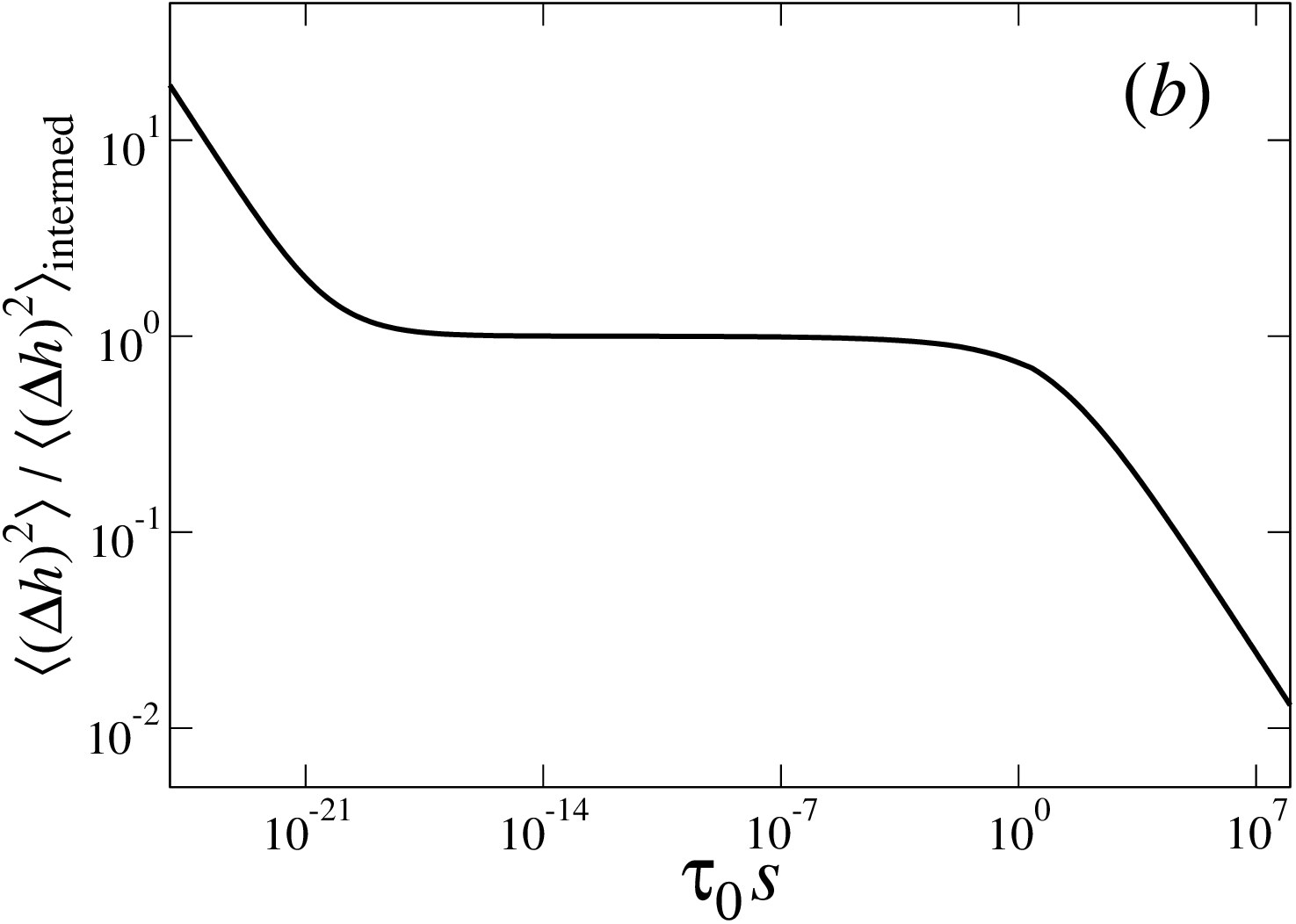}}
\caption{Normalized MSD of a membrane in a semiflexible polymer
  network as a function of frequency $s=i\omega$. (a) The MSD is
  normalized by its solvent-dominated asymptote, $\langle\Delta
  h^2\rangle_{\rm solvent} \sim s^{-5/3}$
  [Eq.~(\ref{MSDsolvent})]. (b) The MSD is normalized by its
  intermediate-regime asymptote, $\langle\Delta h^2\rangle_{\rm
    intermed} \sim s^{-7/5}$ [Eq.~(\ref{MSDintermed})]. The frequency
  in both panels is normalized by
  $\tau_0^{-1}=\kT/(\etas\xi^3)$. Parameters: $\kappa/\kT=10$,
  $\Lp/\xi=100$, $L/\xi=200$.}
\label{fig_MSDsemiflex}
\end{figure}

Figure~\ref{fig_MSDsemiflex} shows the results for the membrane's MSD,
obtained by substituting Eq.~(\ref{Gsemiflex}) in
Eq.~(\ref{MSDgeneral}). The three limiting
behaviors\,---\,solvent-dominated, intermediate, and bulk\,---\,are
clearly seen, and the asymptotic power laws are again confirmed.  As
in the case of flexible polymers, we do not expect the extremely wide
intermediate regime seen in Fig.~\ref{fig_MSDsemiflex} to be
realistic; see Sec.~\ref{sec_discuss}.

\section{Discussion}
\label{sec_discuss}

In this paper we have considered two limiting strengths of coupling
between the membrane and polymer network. The weak-coupling limit
corresponds to a polymer network that does not move together with the
membrane but is merely dragged by the resulting solvent flow. In this
limit we have discovered a new intermediate wavelength regime for the
dispersion relation of membrane undulations, which is translated into
an intermediate regime in time of the dynamics of membrane
roughness. We have found that for both flexible and semiflexible
semidilute polymer solutions this intermediate regime spans several
orders of magnitude in time, until the bulk viscoelasticity takes
over. Yet, it is likely that an earlier crossover should occur, upon
which the weak-coupling limit is no longer valid.

To examine this issue, we first consider the thickness $d$ of the
depletion layer between the membrane and the polymer network. For
both flexible and semiflexible polymers the network exerts on the
membrane an osmotic pressure, $\Pi_{\rm osm}\sim\kT/\xi^3$
\cite{deGennesBook}. The membrane exerts back on the network the
well-known Helfrich repulsion \cite{SafranBook}, $\Pi_{\rm
  Helf}\sim(\kT)^2/(\kappa d^3)$. At equilibrium the two pressures
balance each other, leading to $d\sim (\kT/\kappa)^{1/3}\xi$. We
assume that the weak-coupling limit is valid so long as the membrane's
transverse MSD is smaller than $d^2$. When the MSD is larger, the
polymer directly interacts with the membrane and a crossover to a
different dynamics should take place. Equating Eq.~(\ref{MSDscaling})
with $d^2$, we infer the crossover time,
\[
  t^* \sim \left(\frac{\kappa}{\kT}\right)^{5/6} \tau_\xi,
\]
where $\tau_\xi\sim\etas\xi^3/\kappa$. This limits the intermediate
$t^{2/5}$ behavior of the MSD to $\tau_{\xi}<t<t^*$, suggesting that
only for rigid membranes, where $\kappa \sim 20$--$30$ $\kT$, should
the intermediate regime be observed over a decade in time.

An interesting result can be obtained for a tense membrane or a
surfactant-adsorbed oil-water interface in the weak-coupling limit. In
the intermediate wavelength regime $\lc^{-1}\ll
k\ll(\xi^{-1},\sqrt{\gamma/\kappa})$, where $\gamma$ is the surface
tension, we have $\Omega\simeq \Lambda(k) \gamma k^2\simeq
(\gamma\xi^2/\etas) k^3$. Therefore, the membrane (or interface)
undulation relaxation rate is equivalent to that of a tensionless
membrane in a purely viscous and structureless fluid, with an
effective bending rigidity, $\kappa_{\rm eff}=\gamma\xi^2$. In
experiments in which the relaxation rate of individual modes is aimed
to be measured directly \cite{Tsai2011}, this could make it hard to
determine the physics behind an observation $\Omega(k)\sim k^3$
without a counter measurement of the mean square amplitude of
thermally excited modes (which would be $\approx k_B T/(\gamma
k^2)$). For example, one could (wrongly) interpret such an observation
as indicating membrane stiffening due to polymer adsorption. Yet, the
relaxation rate of the dynamic structure factor is controlled by the
(membrane segment) transverse MSD -- which grows logarithmically with
time, as for a membrane under tension in pure solvent
\cite{Granek1997} -- suggesting that scattering experiments
\cite{Richter2002} could provide an efficient tool even in such
delicate situations. From a different viewpoint, it has been argued
that, for tensed membranes at long lengthscales, a crossover occurs to
dynamics controlled by inter-monolayer friction, which could
complicate further experimental interpretation \cite{Fournier}.

Back to tensionless membranes in the weak-coupling limit, given that
the intermediate regime is observed, our key predictions concerning
its features are as follows: (a) a dispersion relation $\Omega\sim
k^5$; (b) $\mbox{MSD}\sim t^{2/5}$; (c) dynamic structure factor
$S(q,t) \sim \exp[-(\Gamma_q t)^{\beta}]$ and relaxation rate
$\Gamma_q\sim q^z$, with $\beta=2/5$ and $z=5$.  To the best of our
knowledge these universal predictions are yet to be observed in
experiment \cite{Richter2002,polymer-membrane-DLS,Tsai2011}. Neutron
spin echo measurements on a polymer-doped lamellar phase
\cite{Richter2002} showed a decrease of $\beta$ and increase of $z$
from their solvent-dominated values $2/3$ and $3$, respectively, in a
limited $q$ range (see Fig.~10 in Ref.~\cite{Richter2002}). These
results are suggestive but cannot be considered as a validation of our
predictions. Other experimental systems such as polymer-filled
liposomes \cite{Tsai2011} could be used to check the predictions.

\begin{acknowledgement}
This article is dedicated to the dear memory of Lo\"ic Auvray, a
colleague and friend whose wisdom was matched only by his kindness.

\vspace{6pt}
\noindent We are grateful to the Telluride Science Research Center for its
hospitality during a workshop where this work was initiated. We thank
Shigeyuki Komura, Yael Roichman, Pierre Sens, and Matthiew Turner
for helpful comments. HD has been supported by the Israel Science
Foundation (Grant No.~164/14).

\end{acknowledgement}

\section*{Appendix: Boundary-value formulation}
\setcounter{equation}{0}
\renewcommand\theequation{A.\arabic{equation}}

The problem of membrane dynamics inside a two-fluid medium can be
formulated as a boundary-value problem for the medium. Apart from
reproducing the results of the hydrodynamic-kernel approach in the
limits of weak and strong coupling (Sec.~\ref{sec_kernel}), it allows
the treatment of more elaborate boundary conditions, such as a network
that slips at the membrane surface.

For simplicity we assume that the system is uniform along the $y$
axis.  The remaining variables are $(x,z,t)$, turning, after applying
a Fourier transform in $x$ and a Fourier-Laplace transform in $t$,
into $(k,z,\omega)$. The model then contains five $z$-dependent
fields: the velocity $(v_x,v_z)$ and pressure $p$ of the solvent, and
the displacement $(u_x,u_z)$ of the polymer network.

The stresses in the two components (solvent and polymer) are given by
\begin{eqnarray}
  \sigma_{xx}^{\rm s} &=& -p + 2i\etas kv_x \nonumber\\
  \sigma_{xz}^{\rm s} &=& \etas(\pd_zv_x + ikv_z) \nonumber\\
  \sigma_{zz}^{\rm s} &=& -p + 2\etas\pd_zv_z \nonumber\\
  \sigma_{xx}^{\rm p} &=& i(\Kp+4\Gp/3)ku_x + (\Kp-2\Gp/3)\pd_zu_z  \nonumber\\
  \sigma_{xz}^{\rm p} &=& \Gp(\pd_zu_x + iku_x) \nonumber\\
  \sigma_{zz}^{\rm p} &=& (\Kp+4\Gp/3)\pd_zu_z + i(\Kp-2\Gp/3)ku_x.\ \ \
\end{eqnarray}
The five equations for the five fields consist of four force-balance equations,
\begin{eqnarray}
  0 &=& ik\sigma_{xx}^{\rm s} + \pd_z\sigma_{xz}^{\rm s} -\Gamma(v_x-i\omega u_x)
  \nonumber \\
  0 &=& ik\sigma_{xz}^{\rm s} + \pd_z\sigma_{zz}^{\rm s} -\Gamma(v_z-i\omega u_z)
  \nonumber\\
  0 &=& ik\sigma_{xx}^{\rm p} + \pd_z\sigma_{xz}^{\rm p} -\Gamma(i\omega u_x-v_x)
  \nonumber\\
  0 &=& ik\sigma_{xz}^{\rm p} + \pd_z\sigma_{zz}^{\rm p} -\Gamma(i\omega u_z-v_z),
\end{eqnarray}
and an incompressibility condition for the solvent (assuming a dilute
polymer network),
\begin{equation}
  0 = ikv_x + \pd_zv_z.
\end{equation}

Let us count the number of free coefficients to be matched by the
number of boundary conditions. We divide each of the five fields into
the two half-spaces on the two sides of the membrane, $z<0$ and $z>0$,
yielding ten $z$-dependent functions. The equations for the four $v$'s
and four $u$'s are second-order, and the ones for the two $p$'s are
first-order. We need, therefore, 18 boundary conditions. Ten are
provided by demanding that all fields vanish at
$z\rightarrow\pm\infty$. We are left with eight boundary conditions to
be imposed at $z=0$ (the membrane surface). As usual, once these
boundary conditions are stated, one obtains eight linear equations for
the eight coefficients and, demanding the existence of a nontrivial
solution, sets the secular determinant to zero. The resulting
equation, linear in $\omega$, is readily solved for $\omega$ to yield
the dispersion relation, $\Omega=-i\omega$.

We choose to impose in all cases the following six boundary conditions
(the $\pm$ superscripts denote $z\rightarrow 0^\pm$, respectively, and
$\boldsymbol\sigma=\boldsymbol\sigma^{\rm s}+\boldsymbol\sigma^{\rm
  p}$ is the total stress):
\begin{itemize}

\item[(1)] $0=v_x^+ - v_x^-$, continuity of tangential solvent velocity.

\item[(2)] $0=v_z^+ - v_z^-$, continuity of normal solvent
  velocity. The solvent sticks to the membrane; hence, $v_z^\pm$ is
  also equal to the membrane's normal velocity, $i\omega h$.

\item[(3)] $0=u_x^+ - u_x^-$, continuity of tangential polymer displacement.

\item[(4)] $0=u_z^+ - u_z^-$, continuity of normal polymer displacement.

\item[(5)] $0=\sigma_{xz}^+ - \sigma_{xz}^-$, continuity of tangential stress.

\item[(6)] $0=\sigma_{zz}^+ - \sigma_{zz}^- - \kappa k^4 v_z^-$,
  balance of normal forces.

\end{itemize}
The remaining two boundary conditions vary according to the boundary
conditions for the polymer network.

\vspace{0.3cm} {\it Free network}:~~In this case the tangential and
normal stresses exerted by the network both vanish. The two additional
boundary conditions are, therefore,
\begin{itemize}
  \item[(7)] $0=\sigma_{xz}^{\rm p-} - \sigma_{xz}^{\rm p+}$, implying
    that the stresses entering condition (5) are exclusively the solvent's.
  \item[(8)] $0=\sigma_{zz}^{\rm p-} - \sigma_{zz}^{\rm p+}$, implying
    that the stresses entering condition (6), balancing the membrane's
    bending force, are exclusively the solvent's.
\end{itemize}
The resulting dispersion relation is
$\Omega(k,\omega)=\kappa k^4\bar\Lambda(k,\omega)$, where
$\bar\Lambda$ is the kernel obtained in Sec.~\ref{sec_kernel} for the
weak-coupling limit, Eq.~(\ref{Lambdass}).

\vspace{0.3cm} {\it Sticking network}:~~In this case the network
velocity at the membrane is equal to the solvent velocity there (and
their normal components are both equal to the membrane's normal
velocity). The boundary conditions are
\begin{itemize}
  \item[(7)] $0=v_x^- - i\omega u_x^-$, and the same holds for the '+'
    side due to conditions (1) and (3).
  \item[(8)] $0=v_z^- - i\omega u_z^-$, and the same holds for the '+'
      side due to conditions (2) and (4).
\end{itemize}
The resulting dispersion relation is
$\Omega(k,\omega)=\kappa k^4\bar\Lambda(k,\omega)$, where
$\bar\Lambda$ is the kernel obtained in Sec.~\ref{sec_strong} for the
strong-coupling limit, Eq.~(\ref{Lambdastrong}).

\vspace{0.3cm} {\it Slipping network}:~~In this case the normal
component of the network velocity at the membrane is equal to the
solvent's (and both are equal to the membrane's), while the tangential
stress exerted by the network vanishes,
\begin{itemize}
\item[(7)] $0=v_z^- - i\omega u_z^-$, and the same holds for the '+'
  side due to conditions (2) and (4).
\item[(8)] $0=\sigma_{xz}^{\rm p-} - \sigma_{xz}^{\rm p+}$, implying
  that the tangential stresses entering condition (5) are exclusively
  the solvent's.
\end{itemize}
The resulting dispersion relation is
\begin{eqnarray}
  \Omega(k,\omega) = && \frac{\kappa k^3}{4\eta(\omega)} \times \\
  &&\left[1 - 2(\eta/\etas-1) k^2\xi^2 \left(
      1 - \lambda k/\sqrt{1+\lambda^2k^2} \right)\right]^{-1},\nonumber
\end{eqnarray}
which is always positive. This result could not be obtained by the
kernel approach used in the main text. This is because the slip
condition does not set a condition for the relative velocities of the
two components at the membrane, and, thus, cannot be used to determine
the partition of force between the components, as has been done in
Sec.~\ref{sec_kernel}.

Clearly, other choices of boundary conditions can be similarly
studied, such as partial slip between membrane and network, or
asymmetric conditions (\eg network sticks to the membrane on one side
and free on the other).



\end{document}